\documentclass[reprint,
superscriptaddress,
 amsmath,amssymb,
 aps,
 twocolumn,
 longbibliography,
]{revtex4-2}

\usepackage{graphicx,setspace}
\usepackage{dcolumn}
\usepackage{bm}
\usepackage[mathlines]{lineno}
\usepackage[colorlinks=true,citecolor=teal,linkcolor=olive,anchorcolor=green,urlcolor=teal]{hyperref}
\usepackage{mathrsfs}
\usepackage{xcolor}
\usepackage{float}
\usepackage[normalem]{ulem}
\usepackage[utf8]{inputenc}
\usepackage{tensor}
\usepackage{physics}
\usepackage{blindtext}
\usepackage{slashed}
\usepackage{braket}

\newcommand{\be}{\begin{equation}}
\newcommand{\ee}{\end{equation}} 
\newcommand{\ba}{\begin{eqnarray}}
\newcommand{\ea}{\end{eqnarray}}

\begin{document}

\title{Scalar-hairy Lovelock gravity respects zeroth law}
\author{Chaoxi Fang}
\email{chaoxi.f@jxnu.edu.cn}
\affiliation{Department of Physics, Jiangxi Normal University, Nanchang 330022, China}
\author{Libo Xie}
\email{liboxie@jxnu.edu.cn}
\affiliation{Department of Physics, Jiangxi Normal University, Nanchang 330022, China}
\author{Jie Jiang}
\email{jiejiang@mail.bnu.edu.cn}
\affiliation{College of Education for the Future, Beijing Normal University, Zhuhai 519087, China}
\author{Ming Zhang}
\email{mingzhang@jxnu.edu.cn (corresponding author)}
\affiliation{Department of Physics, Jiangxi Normal University, Nanchang 330022, China}

\begin{abstract}
We study the zeroth law for Killing horizon in scalar-hairy Lovelock gravity, and show that the surface gravity of a general Killing horizon in the scalar-hairy Lovelock gravity is constant, provided that the dominant energy condition is obeyed by the matter field. 
\end{abstract}

\maketitle

\section{Introduction}
As one of the keys to open the door of the quantum gravity, the investigation of black hole thermodynamics/mechanics have been stimulating quite a lot of attention since the seminal work \cite{Bardeen:1973gs,Hawking:1976de,Bekenstein:1973ur} by Hawking et al., see Refs. \cite{Kubiznak:2016qmn,Cong:2021fnf} for recent progresses. There is a well correspondence between black hole mechanics and black hole thermodynamics. Remarkably, the mechanical zeroth law of the black hole asserts that the surface gravity of a stationary black hole is constant over the event horizon which is also the Killing horizon. Correspondingly, the thermodynamical zeroth law of the black hole indicates that the temperature of a stationary black hole over the event horizon is invariable.

If the dominant energy condition is respected and the $t-\phi$ orthogonality property is satisfied, the zeroth law was proved in Ref.  \cite{Bardeen:1973gs} for general relativity (GR). Furthermore, in Ref. \cite{Racz:1992bp,Kay:1988mu,Racz:1995nh}, independent of the field equation, it was shown that the surface gravity is constant on the event horizon of a static or stationary-axisymmetric black hole with both  `$t$ - $\phi$' and matter fields reflection isometries. Based on the rigidity theorem \cite{Hawking:1971vc}, the referred event  horizon is also a Killing horizon. Recent progress on the proofs of the zeroth law in the modified gravitational theories can be seen in Refs. \cite{Sarkar:2012wy, Ghosh:2020dkk} for the Lanczos-Lovelock gravity, the scalar-tensor gravity \cite{Dey:2021rke}, and the Horndeski gravity \cite{Sang:2021rla}.  However, a general proof of the black hole zeroth law without any additional condition is still absent, though in Ref. \cite{Racz:1995nh} it was claimed that the constancy of the surface gravity on a Killing horizon is equivalent to vanishing of  the exterior derivative of the twist for the horizon Killing field.

In this compact paper, we aim to study the zeroth law for the Killing horizon in the diffeomorphism invariant scalar-hairy Lovelock gravity \cite{Oliva:2011np}, which contains a real scalar field $\phi$ that is conformally coupled to the Lovelock gravity while the field equations keep to be second order for the metric as well as  for the scalar field. Our motivation is as follows. It was shown in Refs. \cite{Giribet:2014bva,Giribet:2014fla,Galante:2015voa} that the theory with a negative, positive, or vanishing cosmological constant admits black hole solutions with the scalar field being regular everywhere outside the event horizon. Moreover, the AdS case is related to the holographic superconductors \cite{Hartnoll:2008vx}. In Ref. \cite{Hennigar:2016ekz}, the thermodynamic first law of spherically symmetric black hole solution in the theory was constructed; meanwhile, in Ref. \cite{Jiang:2020huc}, the thermodynamic second law of the same black hole solution was proved in the first-order approximation when the perturbation matter fields satisfy the null energy condition. However, all these depend on that the  surface gravity is assumed to be constant, which was not proved yet.

We organize the paper as follows. In Sec. \ref{sec2}, we will review the geometry properties of a stationary Killing horizon. In Sec. \ref{sec3}, we will show the equations of motion for the scalar-hairy Lovelock gravity. The constancy of the surface gravity will be proved in Sec. \ref{sec4}. The last section will be devoted to our conclusion.

\section{GEOMETRY OF STATIONARY KILLING HORIZON}\label{sec2}
For a start, let us review preliminary knowledge about the geometric properties of the Killing horizon (also the event horizon) in a stationary spacetime. A Killing horizon as a null hypersurface $\mathscr{H}$ corresponds to a Killing vector field $\xi^a$ which is timelike except on the horizon where it is null. The Killing horizon yields a function, known as the surface gravity $\kappa$, defined by 
\begin{equation}\label{eqsg}
 \xi^{a}\xi_{; a}^{b}=\kappa \xi^{b}.
\end{equation}
After defining a Killing parameter $v$ by $\xi^a\nabla_a v=1$, the surface gravity can be viewed as the deviation of $v$ with respect to the affine parameter $\lambda$ along the horizon's null geodesic generators.

Checking the zeroth law in the gravity theory is equivalent to testifying whether $\kappa$ is invariable along the hypersurface $\mathcal{H}$. On the Killing horizon, we can construct a basis $\left\{\xi^{a}, l^{a}, e_{i}^{a}\right\}$, where $l^a$ is a null vector satisfying $\xi^{a} l_{a}=-1$, $e_{i}^{a}$ are the tangent vectors on the transverse spacelike surface with induced metric $\gamma_{ab}=\gamma_{i j}\left(e^{i}\right)_{a}\left(e^{j}\right)_{b}$, which lead
\begin{equation}\label{orth}
\begin{gathered}
\xi^{a}\left(e^{i}\right)_{a}=l^{a}\left(e^{i}\right)_{a}=0
\end{gathered}
\end{equation}
on the hypersurface $\mathcal{H}$. With the basis, the transverse metric is given by
\begin{equation}
\gamma_{a b}=2\xi_{(a} l_{b)}+g_{a b},
\end{equation}
with the first term the metric of a two-dimensional hypersurface being orthogonal to the transverse spacelike surface. Besides, the surface gravity $\kappa$ can be written explicitly with the help of the basis according to Eq. (\ref{eqsg}),
\begin{equation}\label{sgk}
\kappa=l_{b} \xi^{a} \nabla_{a} \xi^{b}.
\end{equation}
For a stationary spacetime background, it is evident that 
\begin{equation}
\xi^{a} \nabla_{a} \kappa=0
\end{equation}
for the surface gravity along the geodesic generators. For a stationary background on the Killing horizon, corresponding to the null generator $\xi^a$, on the one hand, we have vanishing expansion scalar which denotes how geodesics defocus or focus; on the other hand, we have vanishing shear, which depicts, for instance, how the geodesic's circular configurations change shape \cite{Blanchette:2021jcw}. Consequently, we arrive at a set of Gauss–Codazzi equations expressing how the Riemann tensor $R_{abcd}$ projects onto the null hypersurface characterized by the basis $\left\{\xi^{a}, l^{a}, e_{i}^{a}\right\}$ that are universal for all gravitational theories,
\begin{equation}\label{gaco}
\begin{aligned}
&R_{a b c d} \xi^{a}\left(e_{i}\right)^{b} \xi^{c}\left(e_{j}\right)^{d}=0, \\
&R_{a b c d} \xi^{a}\left(e_{i}\right)^{b}\left(e_{j}\right)^{c}\left(e_{k}\right)^{d}=0, \\
&R_{a b} \xi^{a} \xi^{b}=0,
\end{aligned}
\end{equation}
where $R_{ab}$ is the Ricci tensor.

The issue of the surface gravity being constant is mathematically equivalent to 
\begin{equation}
\gamma_{a}^{b} \nabla_{b} \kappa=0.
\end{equation}
Substituting Eq. (\ref{sgk}) into it, we have
\begin{equation}
\gamma_{b}^{a} \nabla_{a} \kappa=R_{a e c d} \xi^{a} l^{e} \gamma_{b}^{c} \xi^{d}=-R_{a c} \xi^{a} \gamma_{b}^{c}.
\end{equation}

As an example, we show how to prove the constancy of the surface gravity in the Einstein gravity here. The equation of motion for the spacetime background field reads
\begin{equation}\label{hpai}
H_{a b}=R_{a b}-\frac{1}{2} g_{a b} R=8 \pi T_{a b}.
\end{equation}
First we contract the two sides of Eq. (\ref{hpai}) with  a dyad $\xi^a\xi^b$ and obtain
\begin{equation}
\left.T_{a b} \xi^{a} \xi^{b}\right|_{\mathscr{H}}=0
\end{equation}
by employing Eq. (\ref{gaco}) and the property of the Killing vector field $\xi^a$ being as a null vector. It means that the vector $T_{a b} \xi^{a}$ belongs to a linear space expanded by the Killing vector $\xi^{a}$ and the tangent vectors $\left(e_{i}\right)^{a}$ on the hypersurface $\mathscr{H}$. On the other hand,  we have $T_{a b} \xi^{a}$ being either null or timelike with the assumption that the dominant energy condition is satisfied by the matter field. Thus, $T_{a b} \xi^{a}$ is restricted to be parallel to  $\xi_{a}$, which further yields $T_{a c} \xi^{a} \gamma_{b}^{c}=0$. By contracting the two sides of Eq. (\ref{hpai}) with $\xi^{a} \gamma_{b}^{c}$, we have 
\begin{equation}
\gamma_{b}^{a} \nabla_{a} \kappa=-R_{a c} \xi^{a} \gamma_{b}^{c}=0.
\end{equation}
In consequence we get $\gamma_{a}^{b} \nabla_{b} \kappa=0$ for the Einstein gravity and accordingly the zeroth law in the Einstein gravity is shown to be respected.

\section{Scalar-hairy Lovelock gravity}\label{sec3}
To investigate the zeroth law in the scalary-hairy Lovelock gravity, let us first perform a brief review for the gravity theory, whose action reads
\begin{equation}\label{lag}
I=\frac{1}{16 \pi} \int d^{D} x \sqrt{g}\left(\sum_{k=0}^{k_{\max }} \mathcal{L}^{(k)}+\mathcal{L}_{\mathrm{mat}}\right),
\end{equation}
which describes a gravitational theory in arbitrary dimensions $D$ non-minimally conformally coupled to a real scalar field. $\mathcal{L}_{\mathrm{mat}}$ denotes the extra matter field's Lagrangian density. The Lagrangian densities,  which is of $k-$order,  reads
\begin{equation}
\mathcal{L}^{(k)}=\frac{1}{2^{k}} \delta^{(k)}\left(a_{k} R^{(k)}+b_{k} \phi^{D-4 k} S^{(k)}\right),
\end{equation}
with $k_{\max }=[(D-1) / 2]$, where the brackets signifies to hold the integral part of $(D-1) / 2$ only. Besides,
\begin{equation}
\delta^{(k)}=(2 k) ! \delta_{c_{1}}^{\left[a_{1}\right.} \delta_{d_{1}}^{b_{1}} \cdots \delta_{c_{k}}^{a_{k}} \delta_{d_{k}}^{\left.b_{k}\right]}
\end{equation}
is the generalized Kronecker delta tensor, defined as
\begin{equation}
\delta_{\nu_{1} \nu_{2} \ldots \nu_{p}}^{\mu_{1} \mu_{2} \ldots \mu_{p}}=\operatorname{det}\left(\begin{array}{cccc}
\delta_{\nu_{1}}^{\mu_{1}} & \delta_{\nu_{2}}^{\mu_{1}} & \cdots & \delta_{\nu_{p}}^{\mu_{1}} \\
\delta_{\nu_{1}}^{\mu_{2}} & \delta_{\nu_{2}}^{\mu_{2}} & \cdots & \delta_{\nu_{p}}^{\mu_{2}} \\
\vdots & \vdots & \ddots & \vdots \\
\delta_{\nu_{1}}^{\mu_{p}} & \delta_{\nu_{2}}^{\mu_{p}} & \cdots & \delta_{\nu_{p}}^{\mu_{p}}
\end{array}\right)
\end{equation}
 and the tensors $R^{(k)}$ and $S^{(k)}$ can be explicitly written as
\begin{equation}
R^{(k)}=\prod_{r=1}^{k} R_{a_{r} b_{r}}^{c_{r} d_{r}}, \quad S^{(k)}=\prod_{r=1}^{k} S_{a_{r} b_{r}}^{c_{r} d_{r}},
\end{equation}
with $R_{a b}^{c d}$ the Riemann tensor of the spacetime manifold and
\begin{equation}
\begin{aligned}
S_{a b}^{c d} &=\phi^{2} R_{a b}^{c d}-2 \delta_{[a}^{[c} \delta_{b]}^{d]} \nabla_{e} \phi \nabla^{e} \phi \\
&\quad -4 \phi \delta_{[a}^{[c} \nabla_{b]} \nabla^{d]} \phi+8 \delta_{[a}^{[c} \nabla_{b]} \phi \nabla^{d]} \phi.
\end{aligned}
\end{equation}
Applying the conformal transformation
\begin{equation}
g_{a b} \rightarrow e^{2 \Omega} g_{a b}, \quad \phi \rightarrow e^{s \Omega} \phi,
\end{equation}
we have
\begin{equation}
S_{i j}^{k l} \rightarrow e^{2(s-1) \Omega} S_{i j}^{k l}.
\end{equation}
Specifically, for $s=-1$, we have $\int d^Dx \sqrt{g} \phi^{D-4k}S^{(k)}$ being conformally invariant.

By standard procedure, we can individually derive the equations of motion for the gravitational field and the scalar field,
\begin{equation}\label{eos}
\begin{aligned}
&H_{a b}-H_{a b}^{\phi}=8 \pi T_{a b}, \\
&\sum_{k=0}^{k_{\max }} \frac{(D-2 k) b_{k}}{2^{k}} \phi^{D-4 k-1} \delta^{(k)} S^{(k)}=0,
\end{aligned}
\end{equation}
where
\begin{equation}
H_{a}^{b}=-\sum_{k=0}^{k_{\max }} \frac{a_{k}}{2^{k+1}} \delta_{a c_{1} d_{1} \cdots c_{k} d_{k}}^{b a_{1} b_{1} \cdots a_{k} b_{k}} R_{a_{1} b_{1}}^{c_{1} d_{1}} \cdots R_{a_{k} b_{k}}^{c_{k} d_{k}}
\end{equation}
is the generalized Einstein tensor and $H_{a b}^{\phi}$ is the stress-energy tensor of the scalar field which is explicitly written as
\begin{equation}
\left(H^{\phi}\right)_{a}^{b}=\sum_{k=0}^{k_{\max }} \phi^{D-4 k} \frac{b_{k}}{2^{k+1}} \delta_{a c_{1} d_{1} \cdots c_{k} d_{k}}^{b a_{1} b_{1} \cdots a_{k} b_{k}} S_{a_{1} b_{1}}^{c_{1} d_{1}} \cdots S_{a_{k} b_{k}}^{c_{k} d_{k}}.
\end{equation}
$T_{ab}$ is the stress-energy of the exterior matter fields whose source is denoted as $\varphi$. It was shown in Refs. \cite{Giribet:2014bva,Giribet:2014fla,Galante:2015voa} that the scalary-hairy Lovelock gravity is endowed with some black hole solution together with regularly nonvanishing scalar field outside the horizon.

\section{THE BLACK HOLE ZEROTH LAW IN SCALAR-HAIRY LOVELOCK GRAVITY}\label{sec4}
In what follows, we will explore the zeroth law for the scalar-hairy Lovelock gravity. To this end, we first prove  Gauss-Codazzi-like equations for the tensor $S_{abcd}$. Then we individually contract the two sides of the equation of motion (\ref{eos}) with the tensors $\xi^a \xi^b$ and $\xi^a \gamma^b_c$. Lastly, we show that the zeroth law for the scalar-hairy Lovelock gravity is respected.

\subsection{Gauss-Codazzi-like equations}
We calculate the projection of the tensor $S_{abcd}$ onto the null hypersurface characterized by the basis $\left\{\xi^{a}, l^{a}, e_{i}^{a}\right\}$. First, we contract $S_{abcd}$ with $\xi^{a}\left(e_{i}\right)^{b} \xi^{c}\left(e_{j}\right)^{d}$, which reads
\begin{equation}
\begin{aligned}
&S_{a b}^{c d} \xi^{a}\left(e_{i}\right)^{b} \xi_{c}\left(e^{j}\right)_{d}\\
&=\phi^{2} R_{a b}^{c d} \xi^{a}\left(e_{i}\right)^{b} \xi_{c}\left(e^{j}\right)_{d}\\
&\quad -2 \delta_{[a}^{[c} \delta_{b]}^{d]} \nabla_{e} \phi \nabla^{e} \phi \xi^{a}\left(e_{i}\right)^{b} \xi_{c}\left(e^{j}\right)_{d}\\&\quad-4 \phi \delta_{[a}^{[c} \nabla_{b]} \nabla^{d]} \phi \xi^{a}\left(e_{i}\right)^{b} \xi_{c}\left(e^{j}\right)_{d}\\
&\quad+8 \delta_{[a}^{[c} \nabla_{b]} \phi \nabla^{d]} \phi \xi^{a}\left(e_{i}\right)^{b} \xi_{c}\left(\mathrm{e}^{j}\right)_{d}.
\end{aligned}
\end{equation}
According to the properties of the null vectors $\xi^a$ and $l^a$, together with Eq. (\ref{orth}) and the Gauss-Codazzi equation for the Riemann tensor $R_{abcd}$, it is evident that
\begin{equation}\label{seq1}
S_{a b}^{c d} \xi^{a}\left(e_{i}\right)^{b} \xi_{c}\left(e^{j}\right)_{d}=0.
\end{equation}
Likewise, we have another two projections for the tensor $S_{abcd}$, which are
\begin{equation}\label{seq2}
\begin{aligned}
&S_{a b}^{c d} \xi^{a}\left(e_{i}\right)^{b}\left(e^{j}\right)_{c}\left(e^{k}\right)_{d} \\
&=\phi^{2} R_{a b}^{c d} \xi^{a}\left(e_{i}\right)^{b}\left(e^{j}\right)_{c}\left(e^{k}\right)_{d}\\&\quad-2 \delta_{[a}^{[c} \delta_{b]}^{d]} \nabla_{e} \phi \nabla^{e} \phi \xi^{a}\left(e_{i}\right)^{b}\left(e^{j}\right)_{c}\left(e^{k}\right)_{d}\\&\quad -4 \phi \delta_{[a}^{[c} \nabla_{b]} \nabla^{d]} \phi \xi^{a}\left(e_{i}\right)^{b}\left(e^{j}\right)_{c}\left(e^{k}\right)_{d}\\&\quad +8 \delta_{[a}^{[c} \nabla_{b]} \phi \nabla^{d]} \phi \xi^{a}\left(e_{i}\right)^{b}\left(e^{j}\right)_{c}\left(e^{k}\right)_{d}\\&=0,
\end{aligned}
\end{equation}
and
\begin{equation}\label{seq3}
\begin{aligned}
S_{a d}^{c d} \xi^{a} \xi_{c} &=\phi^{2} R_{a d}^{c d} \xi^{a} \xi_{c}-2 \delta_{[a}^{[c} \delta_{d]}^{d]} \nabla_{e} \phi \nabla^{e} \phi \xi^{a} \xi_{c}\\&\quad-4 \phi \delta_{[\mathrm{a}}^{[\mathrm{c}} \nabla_{\mathrm{d}]} \nabla^{\mathrm{d}]} \phi \xi^{\mathrm{a}} \xi_{\mathrm{c}}+8 \delta_{[\mathrm{a}}^{[\mathrm{c}} \nabla_{\mathrm{d}]} \phi \nabla^{\mathrm{d}]} \phi \xi^{\mathrm{a}} \xi_{\mathrm{c}}\\&=0.
\end{aligned}
\end{equation}
So we can see that the projection properties of the tensor $S_{abcd}$ are similar to those of the Riemann tensor.

\subsection{Contraction with $\xi^{a} \xi^{b}$}
Now let us contract the equation of motion Eq. (\ref{eos}) with the tensor $\xi^{a} \xi^{b}$. By direct calculation and using the properties of the Riemann tensor, we first have
\begin{equation}
\begin{aligned}
&H_{a}^{b} \xi^{a} \xi_{b} \\
&=-\sum_{k=0}^{k_{\max }} \frac{a_{k}}{2^{k+1}} \delta_{a c_{1}d_{1}\cdots c_{k} d_{k}}^{b a_{1} b_{1} \ldots a_{k} b_{k}}   R_{a_{1}b_1}^{c_{1} d_{1}} \cdots R_{a_{k} b_{k}}^{c_{k} d_{k}} \xi^{a} \xi_{b}\\&=\sum_{k=0}^{k_{\max }} \frac{a_{k}}{2^{k+1}} \delta_{a c_{1}d_{1}\cdots c_{k} d_{k}}^{b a_{1} b_{1} \ldots a_{k} b_{k}}   R_{a_{1}b_1}^{c_{1} d_{1}} \cdots R_{a_{k} b_{k}}^{c_{k} d_{k}}  \left(\frac{\partial}{\partial v}\right)^{a}(d u)_b\\&=\sum_{k=0}^{k_{\max }} \frac{a_{k}}{2^{k+1}} \left[\delta^{ui_1j_1\cdots vj_q\cdots i_kj_k}_{vm_1n_1\cdots un_q\cdots m_kn_k} R_{i_{1} j_{1}}^{m_{1} n_{1}} \ldots R_{v j_{q}}^{u n_{q}} \ldots R_{i_{k} j_{k}}^{m_{k} n_{k}}\right.\\&\left.\quad +\delta^{ui_1j_1\cdots vj_q\cdots i_kj_k}_{vm_1n_1\cdots m_qn_q\cdots m_kn_k} R_{i_{1} j_{1}}^{m_{1} n_{1}} \ldots R_{v j_{q}}^{m_q n_{q}} \ldots R_{i_{k} j_{k}}^{m_{k} n_{k}}\right.\\&\left.\quad +\delta^{ui_1j_1\cdots i_qj_q\cdots i_kj_k}_{vm_1n_1\cdots un_q\cdots m_kn_k} R_{i_{1} j_{1}}^{m_{1} n_{1}} \ldots R_{i_q j_{q}}^{u n_{q}} \ldots R_{i_{k} j_{k}}^{m_{k} n_{k}}+\cdots\right]\\&=0,
\end{aligned}
\end{equation}
where in the third step, we have listed three kinds of representative terms and other analogous ones are neglected. Similarly, we can also obtain 
\begin{equation}
\begin{aligned}
&(H^\phi)_{a}^{b} \xi^{a} \xi_{b} \\
&=- \phi^{D-4 k} \sum_{k=0}^{k_{\max }} \frac{b_{k}}{2^{k+1}} \delta_{a c_{1}d_{1}\cdots c_{k} d_{k}}^{b a_{1} b_{1} \ldots a_{k} b_{k}}   S_{a_{1}b_1}^{c_{1} d_{1}} \cdots S_{a_{k} b_{k}}^{c_{k} d_{k}} \xi^{a} \xi_{b}\\&=0.
\end{aligned}
\end{equation}

Thus on the horizon $\mathscr{H}$ we have
\begin{equation}\label{txixi}
8 \pi T_{a b} \xi^{a} \xi^{b}=(H_{a b}-T_{a b}^{\phi}) \xi^{a} \xi^{b}=0.
\end{equation}
Assuming that the dominant energy condition is satisfied by the minimally coupled matter field, then we have $-T_{b}^{a} Z^{b}$ being either timelike or null for a given timelike and future-directed vector $Z^a$. Then once contracting with the null vector $\xi^a$ that is futur-directed, we naturally have $T_{a b} \xi^{a} Z^{b} \geqslant 0$, which yields that $T_{ab}\xi^a$ is either timelike or null. Combing with the fact (\ref{txixi}), we can infer that \begin{equation}\label{emxg}
T_{a b} \xi^{a} \gamma_{c}^{b}=0,
\end{equation}
as $T_{a b} \xi^{b} \propto \xi_{a}$ on the horizon, just like the Einstein case.

\subsection{Contraction with $\xi^{a} \gamma_{c}^{b}$}
To prove the zeroth law in the scalar-hairy Lovelock gravity, we should get  $\gamma_{a}^{b} \nabla_{b} \kappa=0$ for the theory. To this end, what we can use is the equation of motion, for which we contract with the tensor $\xi^{a} \gamma_{c}^{b}$.

First, for the term $H_{ab}$, we have
\begin{equation}
\begin{aligned}
&-H^c_a \xi^{a} \gamma_{bc}\\&=\left(\frac{\partial}{\partial v}\right)^{a} \gamma_{i j}\left(e^{i}\right)_{b}\left(e^{j}\right)_{c}\\&\quad\times \sum_{k=2}^{k_{\max }} \frac{a_{k}}{2^{k+1}} \delta_{a c_{1} d_{1} \cdots c_{k} d_{k}}^{c a_{1} b_{1} \ldots a_{k} b_{k}} R_{a_{1} b_{1}}^{c_{1} d_{1}} \ldots R_{a_{k} b_{k}}^{c_{k} d_{k}}\\&=\gamma_{i j}\left(e^{i}\right)_{b} \sum_{k=2}^{k_{\max }} \frac{a_{k}}{2^{k+1}}\\&\quad \times \left( \delta_{vm_1n_1\cdots j_x u\cdots m_k n_k}^{ji_1 j_1\cdots vu\cdots i_k j_k}R^{m_1 n_1}_{i_1 j_1}\cdots R^{j_x u}_{v u}\cdots R^{m_k n_k}_{i_k j_k}\right.\\&\left.\quad+\delta_{vm_1n_1\cdots j_x n_i\cdots m_nu\cdots m_k n_k}^{ji_1 j_1\cdots vu\cdots i_n j_n\cdots i_k j_k}R^{m_1 n_1}_{i_1 j_1} R^{j_x n_i}_{v u} R^{m_n u}_{i_n j_n}\cdots R^{m_k n_k}_{i_k j_k}\right.\\&\quad+\cdots\\&\quad\left.+\delta_{vm_1n_1\cdots j_x n_i\cdots m_nu\cdots m_k n_k}^{ji_1 j_1\cdots vu\cdots i_n u\cdots i_k j_k}\right.\\&\left.\quad\quad\times R^{m_1 n_1}_{i_1 j_1} R^{j_x n_i}_{v u} R^{m_n u}_{i_n u}\cdots R^{m_k n_k}_{i_k j_k}\right).
\end{aligned}
\end{equation}
According to Eq. (\ref{gaco}), the terms containing $R_{vu}^{j_x n_i}$, $R^{m_n u}_{i_n j_n}$ vanish, leaving
\begin{equation}
\begin{aligned}
&-H^c_a \xi^{a} \gamma_{bc}\\&=\gamma_{i j}\left(e^{i}\right)_{b} \sum_{k=2}^{k_{\max }} \frac{a_{k}}{2^{k+1}}\\&\quad \times \delta_{vm_1n_1\cdots j_x u\cdots m_k n_k}^{ji_1 j_1\cdots vu\cdots i_k j_k}R^{m_1 n_1}_{i_1 j_1}\cdots R^{j_x u}_{v u}\cdots R^{m_k n_k}_{i_k j_k}\\&=-\gamma_{i j}\left(e^{i}\right)_{b} \sum_{k=2}^{k_{\max }} \frac{a_{k}}{2^{k+1}}\\&\quad \times \delta_{j_x m_1n_1\cdots v u\cdots m_k n_k}^{ji_1 j_1\cdots vu\cdots i_k j_k}R^{m_1 n_1}_{i_1 j_1}\cdots R^{j_x u}_{v u}\cdots R^{m_k n_k}_{i_k j_k}\\&=D^{j_x}\kappa \gamma_{i j}\left(e^{i}\right)_{b}  \sum_{k=2}^{k_{\max }} \frac{4k a_{k}}{2^{k+1}} \delta_{j_x m_1n_1\cdots m_k n_k}^{ji_1 j_1\cdots i_k j_k}R^{m_1 n_1}_{i_1 j_1}\cdots R^{m_k n_k}_{i_k j_k}\\&=D^{p}\kappa  \gamma_{ij} \left(e^{i}\right)_{b} M^j_p\\&=\gamma_b^a M^c_a D_c \kappa,
\end{aligned}
\end{equation}
where we have denoted 
\begin{equation}
M_p^j=\sum_{k=2}^{k_{\max }} \frac{4k a_{k}}{2^{k+1}} \delta_{p m_1n_1\cdots m_k n_k}^{ji_1 j_1\cdots i_k j_k}R^{m_1 n_1}_{i_1 j_1}\cdots R^{m_k n_k}_{i_k j_k}.
\end{equation}
Next, we contract the term $(H^\phi)_{a}^{c}$ with  $\xi^{a} \gamma_{bc}$ and obtain 
\begin{equation}\label{hphi1}
\begin{aligned}
&(H^\phi)_{a}^{c}\xi^{a} \gamma_{bc}\\&=-\left(\frac{\partial}{\partial v}\right)^{a} \gamma_{i j}\left(e^{i}\right)_{b}\left(e^{j}\right)_{c}\\&\quad\times \sum_{k=2}^{k_{\max }} \phi ^{D-4 k} \frac{b_{k}}{2^{k+1}} \delta_{a c_{1} d_{1} \cdots c_{k} d_{k}}^{c a_{1} b_{1} \ldots a_{k} b_{k}} S_{a_{1} b_{1}}^{c_{1} d_{1}} \ldots S_{a_{k} b_{k}}^{c_{k} d_{k}}\\&=-\gamma_{i j}\left(e^{i}\right)_{b} \sum_{k=2}^{k_{\max }}\phi ^{D-4 k} \frac{b_{k}}{2^{k+1}}\\&\quad \times \left( \delta_{vm_1n_1\cdots j_x u\cdots m_k n_k}^{ji_1 j_1\cdots vu\cdots i_k j_k}S^{m_1 n_1}_{i_1 j_1}\cdots S^{j_x u}_{v u}\cdots S^{m_k n_k}_{i_k j_k}\right.\\&\left.\quad+\delta_{vm_1n_1\cdots j_x n_i\cdots m_nu\cdots m_k n_k}^{ji_1 j_1\cdots vu\cdots i_n j_n\cdots i_k j_k}S^{m_1 n_1}_{i_1 j_1} S^{j_x n_i}_{v u} S^{m_n u}_{i_n j_n}\cdots S^{m_k n_k}_{i_k j_k}\right.\\&\quad+\cdots\\&\quad\left.+\delta_{vm_1n_1\cdots j_x n_i\cdots m_nu\cdots m_k n_k}^{ji_1 j_1\cdots vu\cdots i_n u\cdots i_k j_k}\right.\\&\left.\quad\quad\times S^{m_1 n_1}_{i_1 j_1} S^{j_x n_i}_{v u} S^{m_n u}_{i_n u}\cdots S^{m_k n_k}_{i_k j_k}\right).
\end{aligned}
\end{equation}
Likewise, by employing Eqs. (\ref{seq1})--(\ref{seq3}), we can reduce Eq. (\ref{hphi1})  as
\begin{equation}
\begin{aligned}
&(H^\phi)_{a}^{c}\xi^{a} \gamma_{bc}\\&=-\gamma_{i j}\left(e^{i}\right)_{b} \sum_{k=2}^{k_{\max }} \phi ^{D-4 k} \frac{b_{k}}{2^{k+1}}\\&\quad \times \delta_{vm_1n_1\cdots j_x u\cdots m_k n_k}^{ji_1 j_1\cdots vu\cdots i_k j_k}S^{m_1 n_1}_{i_1 j_1}\cdots S^{j_x u}_{v u}\cdots S^{m_k n_k}_{i_k j_k}\\&=\gamma_{i j}\left(e^{i}\right)_{b} \sum_{k=2}^{k_{\max }}\phi ^{D-4 k} \frac{b_{k}}{2^{k+1}}\\&\quad \times \delta_{j_x m_1n_1\cdots v u\cdots m_k n_k}^{ji_1 j_1\cdots vu\cdots i_k j_k}S^{m_1 n_1}_{i_1 j_1}\cdots S^{j_x u}_{v u}\cdots S^{m_k n_k}_{i_k j_k}\\&=-D^{j_x}\kappa \sum_{k=2}^{k_{\max }}\phi ^{D-4 k} \frac{4k b_{k}}{2^{k+1}} \delta_{j_x m_1n_1\cdots m_k n_k}^{ji_1 j_1\cdots i_k j_k}S^{m_1 n_1}_{i_1 j_1}\cdots S^{m_k n_k}_{i_k j_k}\\&=-D^{p}\kappa \gamma_{ij} \left(e^{i}\right)_{b} N^j_p\\&=-\gamma_b^a N^c_a D_c \kappa,
\end{aligned}
\end{equation}
where we have denoted 
\begin{equation}
N_p^j=\sum_{k=2}^{k_{\max }}\phi ^{D-4 k}  \frac{4k b_{k}}{2^{k+1}} \delta_{p m_1n_1\cdots m_k n_k}^{ji_1 j_1\cdots i_k j_k}S^{m_1 n_1}_{i_1 j_1}\cdots S^{m_k n_k}_{i_k j_k}.
\end{equation}
Thus, we can obtain the identity
\begin{equation}
\left[H^c_a \xi^{a}+(H^\phi)_{a}^{c}\xi^{a}\right]\gamma_{bc}=\gamma_b^a M^c_a D_c \kappa-\gamma_b^a N^c_a D_c \kappa.
\end{equation}
Combining Eq. (\ref{emxg}), we immediately have
\begin{equation}\label{mndk}
\gamma_b^a M^c_a D_c \kappa-\gamma_b^a N^c_a D_c \kappa=0.
\end{equation}

\subsection{Generalized zeroth law for the scalar-hairy Lovelock gravity}
We can rewrite Eq. (\ref{mndk}) in form of 
\begin{equation}\label{bz49}
Q_a^b Z_b=0,
\end{equation}
where 
\begin{equation}
Q_a^b \equiv \gamma_a^c M^b_c -\gamma_a^c N^b_c,
\end{equation}
\begin{equation}
Z_b \equiv D_b \kappa.
\end{equation}
As all quantities can be reduced to the Einstein case in the limits $a_k\to 0, b_{k^\prime}\to 0$, where $k\geqslant 2, k^\prime \geqslant 0$, we can expand both $Q_b^c$ and $Z_a$ in terms of $\mathbb{A}\to 0$,
\begin{equation}
Q_{a}^{b}=\left(Q_{0}\right)_{a}^{b}+\mathbb{A}\left(Q_{1}\right)_{a}^{b}+\mathbb{A}^{2}\left(Q_{2}\right)_{a}^{b}+\cdots,
\end{equation}
\begin{equation}
Z_{a}=\left(Z_{0}\right)_{a}+\mathbb{A}\left(Z_{1}\right)_{a}+\mathbb{A}^{2}\left(Z_{2}\right)_{a}+\cdots,
\end{equation}
with $(Q_0)_a^b$ and $(Z_0)_a$ the Einstein counterparts for $Q_a^b$ and $Z_a$, respectively. Thus the Eq. (\ref{bz49}) can be reorganized as 
\begin{equation}
\begin{aligned}
0=&\left(Q_{0}\right)^{b}_a\left(Z_{0}\right)_{b}+\mathbb{A}\left[\left(Q_{0}\right)^{b}_a\left(Z_{1}\right)_{b}+\left(Q_{1}\right)^{b}_a\left(Z_{0}\right)_b\right]\\&+\mathbb{A}^{2}\left[\left(Q_{0}\right)_{a}^{b}\left(Z_{2}\right)_{b}+\left(Q_{1}\right)^{b}_a\left(Z_{1}\right)_{b}+\left(Q_{2}\right)^{b}_a\left(Z_{0}\right)_{b}\right]\\&+\cdots.
\end{aligned}
\end{equation}
To make the last identity hold for any $\mathbb{A}$, we just need that every order of $\mathbb{A}$ vanishes on the right side. Since we already have $\left(Z_{0}\right)_{b}=0$, the first order of $\mathbb{A}$ can be reduced as $\left(Q_{0}\right)^{b}_a\left(Z_{1}\right)_{b}+\left(Q_{1}\right)^{b}_a\left(Z_{0}\right)_b=\left(Q_{0}\right)^{b}_a\left(Z_{1}\right)_{b}$. As $\left(Q_{0}\right)^{b}_a\neq 0$, we just get $\left(Z_{1}\right)_{b}=0$. Then we have $\left(Q_{0}\right)_{a}^{b}\left(Z_{2}\right)_{b}+\left(Q_{1}\right)^{b}_a\left(Z_{1}\right)_{b}+\left(Q_{2}\right)^{b}_a\left(Z_{0}\right)_{b}=\left(Q_{0}\right)_{a}^{b}\left(Z_{2}\right)_{b}=0$ for the second order of $\mathbb{A}$, which yields $\left(Z_{2}\right)_{b}=0$. Similarly, we have $\left(Q_{0}\right)_{a}^{b}\left(Z_{k}\right)_{b}=0$ for the $k$th order of $\mathbb{A}$, which indicates that $(Z_k)_a=0$. Eventually, we get $Z_a=0$. Therefore, we prove that the zeroth law is respected by the scalar-hairy Lovelock gravity, with the assumption that the gravity theory can be smoothly reduced to the Einstein gravity.

\section{Conclusion and Discussion}\label{sec5}
The zeroth law in the scalar-hairy Lovelock gravity was investigated. To this end, we first gave a brief review of the proof for the zeroth law in the Einstein gravity. Then applying the equations of motion in the scalar-hairy Lovelock gravity, we  calculated the derivative of the surface gravity $\kappa$ for the stationary black hole in each direction along the event horizon. We obtain an identity Eq. (\ref{mndk}) adapting to the surface gravity along the transverse directions, as long as the dominant energy condition is assumed to be satisfied. Furthermore, on the basis that all the quantities can be reduced to the ones in the Einstein gravity, we showed that 
\begin{equation}
\gamma_{b}^{a} \nabla_{a} \kappa=0
\end{equation}
for the scalar-hairy black hole. Thus, we proved that the surface gravity is constant and the zeroth law is respected.

As pointed out in Ref. \cite{Sarkar:2012wy}, the extension of the zeroth law from the Einstein gravity to the scalar-hairy Lovelock gravity must be assured by two facts. The first step is that the surface gravity for the Killing horizon of the scalar-hairy Lovelock gravity should be proved and the second step is that the rigidity theorem should also be proved for this theory to ensure that the event horizon is also a Killing horizon. We have finished the first proof in this paper and have to leave the second one as an open issue for future work.

Furthermore, the extension of the zeroth law for the scalar-hairy Lovelock gravity from a branch which can reduce to the Einstein gravity smoothly to one that is characterized with $k-$fold degeneration is a subtle issue. For an $N$th  order  Lovelock gravity, there can be  as many as $k$ $(k\leqslant N)$ effective cosmological constants in the field equation taking a  same value, resulting in a $k-$fold degenerate AdS vacuum. Then a critical point  in the parameter space obstructing the theory to be linearized emerges \cite{Fan:2016zfs,Camanho:2013pda}. The generalization of the zeroth law to the so-called Lovelock Unique Vacuum \cite{Arenas-Henriquez:2017xnr,Arenas-Henriquez:2019rph}, which contains Born-Infeld AdS gravity  and Chern-Simons AdS gravity  in the maximal multiplicity case for even dimensions and odd dimensions respectively, is thus one other challenging topic to be studied.

\section*{Acknowledgements}
M. Z. is supported by the National Natural Science Foundation of China (Grant No. 12005080) and Young Talents Foundation of Jiangxi Normal University (Grant No. 12020779). J. J. is supported by the Talents Introduction Foundation of Beijing Normal University (Grant No. 310432102) and  the GuangDong Basic and Applied Basic Research Foundation with Grant No. 217200003.

\end{document}